\documentclass[12pt]{article}
\renewcommand{\baselinestretch}{1.1}
\usepackage{graphicx}
\usepackage{epsfig}
\begin{document}
\topskip 2cm
\renewcommand{\thefootnote}{\fnsymbol{footnote}}  
\begin{titlepage}
\rightline{ \large{ \bf March 2002} }
\begin{center}
{\Large\bf Numerical evaluation of the general massive }\\ 
{\Large\bf 2-loop sunrise self-mass master integrals }\\
 {\Large\bf from differential equations. } \\

\vspace{2.5cm}
{\large {\bf
M.~Caffo$^{ab}$, 
H.~Czy{\.z}\ $^{c}$}}\footnote{Work supported in part by the 
European Community's Human Potential Programme under contract 
HPRN-CT-2000-00149 Physics at Colliders.}
{\large and   
{\bf E.~Remiddi$^{ba}$ \\ } }

\begin{itemize}
\item[$^a$]
             {\sl INFN, Sezione di Bologna, I-40126 Bologna, Italy }
\item[$^b$] 
             {\sl Dipartimento di Fisica, Universit\`a di Bologna, 
             I-40126 Bologna, Italy }
\item[$^c$] 
             {\sl Institute of Physics, University of Silesia, 
             PL-40007 Katowice, Poland }

\end{itemize}
\end{center}

\noindent
e-mail: {\tt caffo@bo.infn.it \\ 
\hspace*{1.3cm} czyz@us.edu.pl \\ 
\hspace*{1.3cm} remiddi@bo.infn.it \\ } 
\vspace{.5cm}
\begin{center}
\begin{abstract}
The system of 4 differential equations in the external invariant satisfied 
by the 4 master integrals of the general massive 2-loop sunrise self-mass 
diagram is solved by the Runge-Kutta method in the complex plane. 
The method, whose features are discussed in details, offers a reliable 
and robust approach to the direct and precise numerical evaluation 
of Feynman graph integrals.  
\end{abstract}
\end{center}
\scriptsize{ \noindent ------------------------------- 
    \\ } 
\vfill
\end{titlepage}
\pagestyle{plain} \pagenumbering{arabic} 
\newcommand{\Eq}[1]{Eq.(\ref{#1})} 
\newcommand{\labbel}[1]{\label{#1}} 
\newcommand{\cita}[1]{\cite{#1}} 
\def\Re{\hbox{Re~}} 
\def\Im{\hbox{Im~}} 
\newcommand{\F}[1]{F_#1(n,m_1^2,m_2^2,m_3^2,p^2)} 
\newcommand{\dnk}[1]{ d^nk_{#1} } 
\newcommand{\Fn}[2]{F_{#1}^{(#2)}(m_1^2,m_2^2,m_3^2,p^2)}
\newcommand{\D}{D(m_1^2,m_2^2,m_3^2,p^2)} 

\section{Introduction.} 

High precision measurements in high energy physics (or more in general
in the determination of particle properties) require more and 
more precise calculations of multi-loop Feynman diagrams to have 
sufficiently precise theoretical predictions to compare with. 

The nowadays widely accepted procedure of expressing
radiative correction calculations in terms of a limited number of 
master integrals (MI) \cite{TkaChet} reduces 
the problem to the careful determination of these quantities. 
The method has also the advantage that, with a correct bookkeeping of 
the recurrence relations arising from integration by parts identities, 
the MI of a given problem can be reused in 
more complicated calculations.

The analytical calculation of MI, in terms of the usual 
polylogarithms and their generalizations, is in general possible 
only when the number of different scales (internal masses 
and external momenta 
or Mandelstam variables) is small, like in QCD calculations, where 
all masses are set to zero or in the QED cases, where only the 
electron mass is different from zero, or when the external variables
are fixed to particular values (zero or mass shell condition). 
Another possibility of big help in analytic calculations is sometimes 
offered by the exploitation of particular simplifying conditions, like the 
smallness of some ratios of the parameters allowing the corresponding 
expansion.  

In the general massive case, relevant in the electroweak theory, 
the number of parameters prevents from obtaining results in 
the usual 
analytic form already in the case of the 2-loop sunrise self-mass 
diagram shown in Fig.1. \\ 

\begin{figure}[!h]
\epsfbox[0 20 140 140]{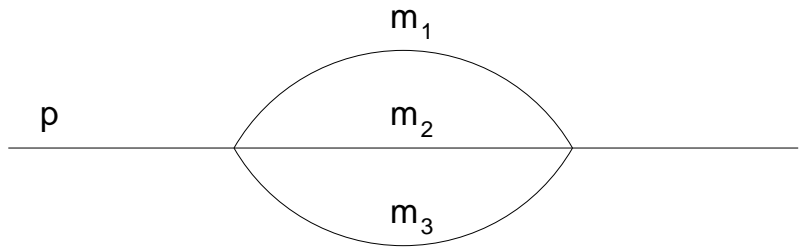}
\caption{The general massive 2-loop sunrise self-mass diagram.} 
\end{figure}

This diagram has indeed a long history of investigation and its MI were 
even recognized to be expressible in closed form as a combination of four 
Lauricella functions, a special class of generalized hypergeometric 
series \cite{BBBS} (and earlier references therein). 
The method provides efficient multiple series expansions for the regions
of small $|p^2|$, i.e. $|p^2| < max(m_i^2)$, and of large 
$|p^2|$, i.e. $|p^2| > (m_1+m_2+m_3)^2$, but some problems arise in the
intermediate region.

Great efforts were therefore devoted to investigate the properties 
in the special points (i.e. $p^2 = 0, \infty$, pseudothresholds 
and threshold).
The analytical expansions of the MI at $0$ and $\infty$ are given in 
\cite{BBBS} and \cite{CCLR1}; the values at pseudothresholds and 
threshold in \cite{BDU};
the analytical expansions at pseudothresholds are in \cite{CCR1};  
a semi-analytical expansion at threshold is in \cite{DS} and also in 
configuration space technique in \cite{GP}, while the 
complete analytical expansions at threshold are presented  
in \cite{CCR2}. 

For numerical evaluation purposes, it is possible to cast 
the general massive self-mass diagram as  
a double integral representation  
and in the particular case of the sunrise diagram in a single integral 
representation \cite{BBBS}, \cite{PT} (and earlier references therein).
The configuration space technique is also exploited in the numerical 
approach \cite{BBBS}, \cite{GKP}.
In a recent approach rearrangements of the integrand, driven by the 
Bernstein-Tkachov theorem, are introduced to improve numerical 
convergence \cite{Passarino}. 
A different and interesting method is the use of the recurrence relations
as difference equations to numerically evaluate the MI \cite{Laporta}.

In the present paper we exploit the numerical evaluation of the four MI 
related to the general massive 2-loop sunrise self-mass diagram 
\cite{Tarasov}, using 
the differential equations in $p^2$, obtained in \cite{CCLR1} and the 
Runge-Kutta method \cite{PTVF} to solve them for complex values of $p^2$.
The interest is not limited to provide with a fast routine precise 
numerical values for all the four MI in the general massive case and for 
all the values of $p^2$, but extends to the investigation of the reliability 
of the method, as it can be easily extended to the numerical 
evaluation of other less studied diagrams. 

In Section 2 the master differential equations are recalled and the 
analytic properties of the MI are reviewed. 
Section 3 contains a description of the method used to solve 
the system of differential equations and to determine the accuracy.  
In Section 4 the control tests and the comparisons with other values 
reported in the literature are presented. 
Finally in Section 5 our conclusions on the application of the method 
to present and further work are presented.


\section{Analytical properties and behaviours of the MI.}

We use here the following definition of the four MI related to the 
general massive 2-loop sunrise self-mass diagram in $n$ continuous 
dimensions and with fully Euclidean variables

\begin{eqnarray} 
 \F{j}
      &=& \frac{ \mu^{8-2n}}{((2\pi)^{n-2})^2 }\nonumber \\  
      && {\kern-120pt} \int \dnk{1} \int \dnk{2} \; 
      \frac{ 1 } 
           { (k_1^2+m_1^2)^{\alpha_1(j)} (k_2^2+m_2^2)^{\alpha_2(j)} 
             ( (p-k_1-k_2)^2+m_3^2 )^{\alpha_3(j)} } \ , j=0,1,2,3 
\labbel{1} \end{eqnarray} 

and $ i=1,2,3; \ \alpha_i(0)=1$, for $j=0$; \(\alpha_i(j)=1\), for \(j\ne i\);
 \(\alpha_i(j)=2\), for \(j = i\).

Wherever necessary to avoid ambiguities, the usual imaginary displacements 
$m_i^2 \to m_i^2 -i \epsilon$, where $\epsilon$ is an 
infinitesimal positive number, are understood.  

At variance from \cite{CCLR1}, were the mass scale was given 
the value \( \mu = 1, \) here we choose 
\begin{equation} 
 \mu = m_1+m_2+m_3 \ , 
\labbel{mu} \end{equation} 
which comes out to be the appropriate 
mass scale parameter for the numerical discussion. 
The expansion of the MI around \( n = 4 \) has the form \cite{CCLR1}
\begin{eqnarray} 
 && \F{j} = C^2(n) \Biggl\{  \frac{1}{(n-4)^2} \Fn{j}{-2} 
            \nonumber \\ 
 && {\kern+50pt} + \frac{1}{(n-4)}   \Fn{j}{-1} 
                  + \Fn{j}{0} + {\cal O} (n-4) \Biggr\} \ . 
\labbel{39a} \end{eqnarray} 
where the coefficient $C(n)$  
\begin{equation} 
C(n) = \left(2 \sqrt{\pi} \right)^{(4-n)} \Gamma\left(3-\frac{n}{2}\right) 
                                                               \ , 
\labbel{16a} \end{equation} 
not expanded, can be replaced by its value $C(4)=1$, at $n = 4$, 
when multiplying a function regular in \( (n-4) \). 
The coefficients of the poles in \( (n-4) \) of \( \F{0} \) are 
known to be \cite{CCLR1}  
\begin{eqnarray} 
 \Fn{0}{-2} &=& -\frac{1}{8} (m_1^2+m_2^2+m_3^2) \ ,
\nonumber \\
 \Fn{0}{-1} &=&  \frac{1}{8} \Biggl\{ \frac{p^2}{4} 
                +\frac{3}{2} (m_1^2+m_2^2+m_3^2) 
\nonumber \\ && 
     - \Biggl[ m_1^2 \log(\frac{m_1^2}{\mu^2}) 
             +m_2^2 \log(\frac{m_2^2}{\mu^2}) 
             +m_3^2 \log(\frac{m_3^2}{\mu^2}) \Biggr] \Biggr\} \ , 
\labbel{39b} \end{eqnarray} 
while those for \( \F{i}, i=1,2,3 \) are 
\begin{eqnarray} 
 \Fn{i}{-2} &=& \frac{1}{8} \ ,
\nonumber \\
 \Fn{i}{-1} &=&  - \frac{1}{16} + \frac{1}{8} \log(\frac{m_i^2}{\mu^2}) \ . 
\labbel{39c} \end{eqnarray} 

From now on we deal only with the finite parts of the MI 
(i.e. we subtract the \(n-4\) poles) 
and we do not write anymore, for short, the arguments of the functions
\(\Fn{j}{0} \equiv F_j^{(0)} \), unless we need to refer explicitly to them. 

The differential equations satisfied by the finite part of the MI 
expansions, given 
in \cite{CCLR1}, can be written as  
\begin{eqnarray} 
p^2 {\frac{\partial} {\partial p^2}} F^{(0)}_0 =
 F^{(0)}_0 + m_1^2F^{(0)}_1+m_2^2F^{(0)}_2+m_3^2F^{(0)}_3 +T_0 \nonumber \\ 
 p^2 \ \D 
  \ {\frac{\partial}{\partial p^2}} F^{(0)}_i =
 \sum_{j=0}^{3} M_{i,j}  F^{(0)}_j + T_i  \ \ , \ i=1,2,3
 \labbel{a6}
 \end{eqnarray} 
\noindent
where the explicit form of the functions \(T_0, T_i\) 
(polynomials of $p^2$ and $m_i^2$, 
and logarithms of $m_i^2/\mu^2$) and \(M_{i,j}\) (polynomials of $p^2$ and 
$m_i^2$) can be found in \cite{CCLR1}.
 The function \(\D\) is defined by
 \begin{eqnarray} 
\D&=&
  \left(p^2-p_{th }^2\right) \left(p^2-p_{ps1}^2\right) 
  \left(p^2-p_{ps2}^2\right) \left(p^2-p_{ps3}^2\right) \ ,\\
  p_{th }^2 &=& -(m_1+m_2+m_3)^2 \ , \nonumber \\
  p_{ps1}^2 &=& -(m_1+m_2-m_3)^2 \ , \nonumber \\ 
  p_{ps2}^2 &=& -(m_1-m_2+m_3)^2 \ , \nonumber \\
  p_{ps3}^2 &=& -(m_1-m_2-m_3)^2 \ , \nonumber 
  \labbel{a7}
 \end{eqnarray} 
\noindent
and vanishes at the threshold $ p_{th }^2 $ and at the 
three pseudothresholds $ p_{ps1}^2, p_{ps2}^2, p_{ps3}^2$. 
Indeed the values of $ p^2 $ at which the coefficients of the derivatives 
in \Eq{a6} vanish, together with $p^2=\infty$, are the singular points 
of the differential equations, which we will call special points because
special care is required in the numerical computation of the MI.  
 
The differential equations \Eq{a6} allow a class of solutions wider 
than just the functions \Eq{1}, but the initial conditions at \(p^2=0\)
imposed by \Eq{1} identify uniquely the solutions 
(actually the regularity at \(p^2=0\), even without the 
explicit knowledge of the functions in that point, is enough to fix 
the solutions). 
Once the initial condition at \(p^2=0\) is fixed, one finds that the
pseudo-threshold values \(p^2 = p_{ps1}^2 \),
\(p^2 = p_{ps2}^2 \), \(p^2 = p_{ps3}^2 \) are also
regular points \cite{CCR1}, while the threshold value \(p^2 = p_{th }^2 \)
is a branch point \cite{CCR2} (in agreement, of course, with 
standard textbook results \cite{IZ}).  

It is convenient to use reduced masses and reduced external invariant 
\begin{equation} 
m_{i,r} \equiv \frac{m_i}{m_1+m_2+m_3} \quad, \quad 
p^2_r \equiv \frac{p^2}{(m_1+m_2+m_3)^2} \ , 
\labbel{red1} \end{equation} 
together with a dimensionless version of \( F^{(0)}_0 \), defined by 
\begin{equation} 
 F^{(0)}_{0,r} \equiv \frac{F^{(0)}_0}{(m_1+m_2+m_3)^2} \ ; 
\labbel{red2} \end{equation} 
as the other master integrals are already dimensionless, the values 
of all the functions are now pure numbers. 
In terms of the new variables \(p^2_r, m_{i,r}\) the threshold is located at  
\( p_{th,r}^2 = p_{th}(m_{i,r}) = -1 \) and the 
pseudo-thresholds are in 
$p_{ps1,r}^2= p_{ps1}(m_{i,r}), p_{ps2,r}^2= p_{ps2}(m_{i,r}), 
 p_{ps3,r}^2= p_{ps3}(m_{i,r})$. 
We also recall here that according to the Euclidean definition our 
\(p^2\) is positive for space-like $p$, and negative for time-like $p$.  

\begin{figure}[ht]
\begin{center}
\epsfig{file=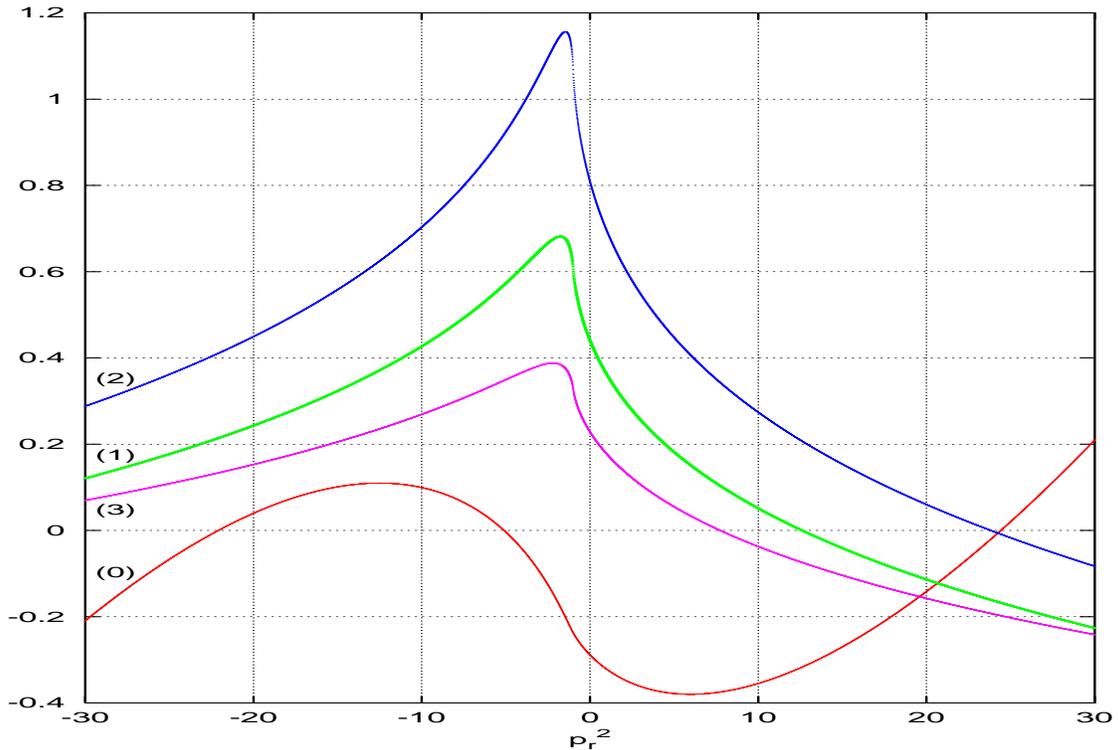,width=\textwidth,height=10cm}
\end{center}
\caption{ Plots of  
 \(\Re F^{(0)}_{0,r}\) (labeled as (0)) and \(\Re F^{(0)}_{i}\) 
(labeled as ($i$)) as a function of \(p_r^2\) for \(m_1 \ = \ 2, \ 
m_2\ = \ 1, \ m_3\ =\ 4 \) and  \(\mu = m_1+m_2+m_3 \). }
\label{fig:f1}
\end{figure}

\begin{figure}[ht]
\begin{center}
\epsfig{file=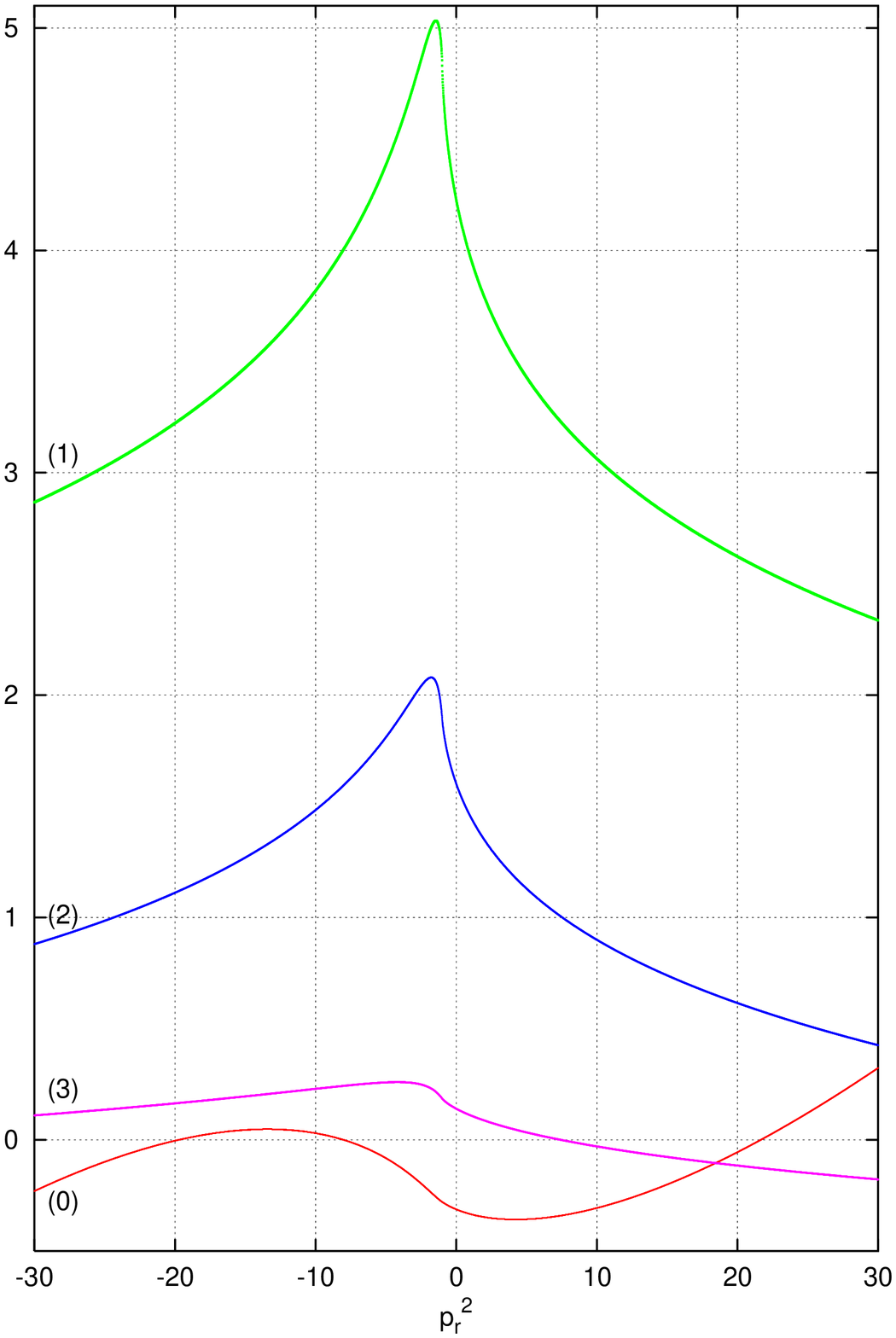,width=\textwidth,height=10cm}
\end{center}
\caption{Plots of 
 \(\Re F^{(0)}_{0,r}\) (labeled as (0)) and \(\Re F^{(0)}_{i}\) 
(labeled as ($i$)) as a function of \(p_r^2\) for \(m_1 \ = \ 1, \
m_2\ = \ 9, \  m_3\ =\ 200 \) and  \(\mu = m_1+m_2+m_3 \). }
\label{fig:f2}
\end{figure}

Typical plots of the real and imaginary parts of the MI
for two sets of masses are shown in Figures 
{\ref{fig:f1}},{\ref{fig:f2}},{\ref{fig:f3}} and {\ref{fig:f4}}. 
The plots are obtained by means of 
the FORTRAN program described in the next two sections and 
each consists of 6000 points per function calculated with the relative 
precision of \(10^{-6}\).

The behavior of the functions depends of course  
on the values of the masses, 
however some of their properties are quite general, as it appears  
from the following discussion.
The real part of \(F_{0,r}^{(0)}\) as seen in  figures {\ref{fig:f1}} 
and {\ref{fig:f2} has one local minimum and one local maximum
for finite $p^2_r$ and goes to \(\pm \infty\) for \( p^2_r \to \pm \infty \). 
We do not have analytically the exact position of the local extrema, 
but as they appear to lie outside the region $ -1 \le p^2_r \le 1 $ 
they can be found approximately with the help of 
the asymptotic expansion (large \( p^2_r\)) for \(F_{0,r}^{(0)}\), 
which is \cite{CCLR1}
\begin{eqnarray}
 F^{(0)}_{0,r} = \frac{p^2_r}{32} \left(\log(p^2_r)-\frac{13}{4}\right)
 +\frac{1}{32} \log^2(p^2_r)\sum_{i=1}^3 m_{i,r}^2
 -\frac{1}{16} \log(p^2_r)\sum_{i=1}^3 m_{i,r}^2\log(m_{i,r}^2)
  + \cdots \ .
\labbel{an1}
\end{eqnarray} 

The asymptotic behavior of \( \Re F_{0,r}^{(0)} \) is obvious from \Eq{an1},
while for the positions of the maximum and minimum one finds approximately
from the first term only, which is also the leading, \( p^2_{r,max} = -9.5\) 
and \( p^2_{r,min} = 9.5\), independent from the mass values. 
The approximate values of the function at those points, 
again from the first 
term only, are \(\Re F^{(0)}_{0,r}(p^2_{r,max}) = 0.3\) and 
\(\Re F^{(0)}_{0,r}(p^2_{r,min}) = -0.3\), also independent from the mass 
values. 
Taking into account the asymptotic behavior one expects at least three zeros 
of the function, provided that in non asymptotic region there are no 
additional extrema (which is actually the case).
The second derivative of \(\Re F_{0,r}^{(0)}\) at threshold is infinite 
\cite{CCR2}, but it does not change sign at that point, even if the position 
of the flex point of \(\Re F_{0,r}^{(0)}\) is not far from the threshold.
 
The other MI go one into the other by the exchange of the values of the 
related masses. 
From the expression of their expansion for large \( p^2_r\)

  \begin{eqnarray}
  F^{(0)}_{i} = -\frac{1}{32} \log^2(p^2_r)
  +\frac{1}{16} \log(p^2_r)\left(\log(m_{i,r}^2)+1\right)+ \cdots \ ,
 \labbel{an2}
 \end{eqnarray} 
\noindent
we see that their real parts all go to \(-\infty\) in both asymptotic regions 
\( p^2_r \to \pm \infty \), however the position of the maximum 
cannot be obtained just from the first terms of the asymptotic expansion, 
as it is positioned in the region of small $p^2_r$.  
The analytic expansions at threshold \cite{CCR2} show that the 
derivatives of the \(\Re F^{(0)}_{i}\) are infinite, but they do not 
change sign exactly at that point.  

\begin{figure}[ht]
\begin{center}
\epsfig{file=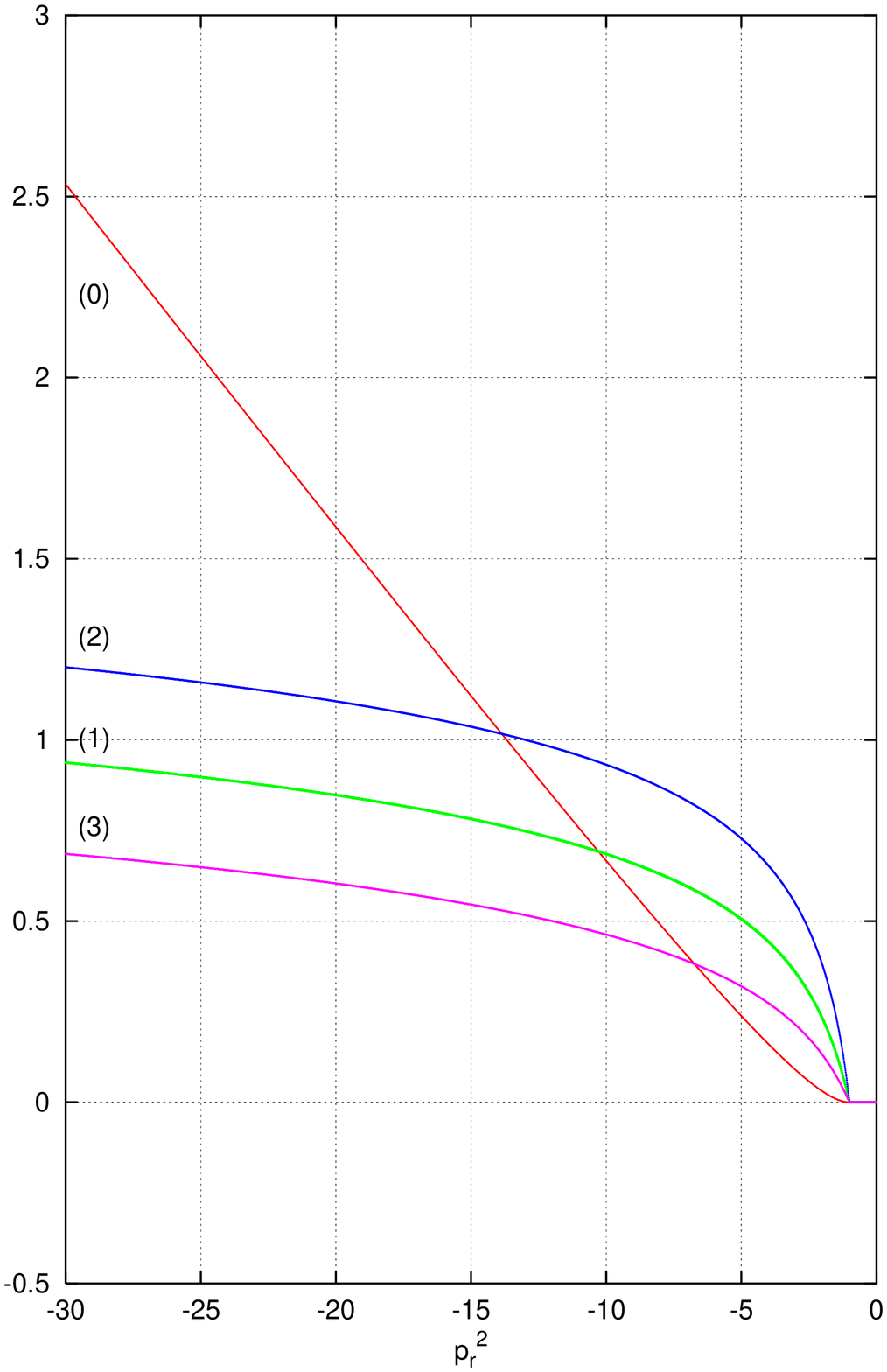,width=\textwidth,height=10cm}
\end{center}
\caption{Plots of 
\(\Im F^{(0)}_{0,r}\) (labeled as (0)) and \(\Im F^{(0)}_{i}\) 
(labeled as ($i$)) as a function of \(p_r^2\) for \(m_1 \ = \ 2, \ 
m_2\ = \ 1, \  m_3\ =\ 4 \) and  \(\mu = m_1+m_2+m_3 \). }
\label{fig:f3}
\end{figure}

\begin{figure}[ht]
\begin{center}
\epsfig{file=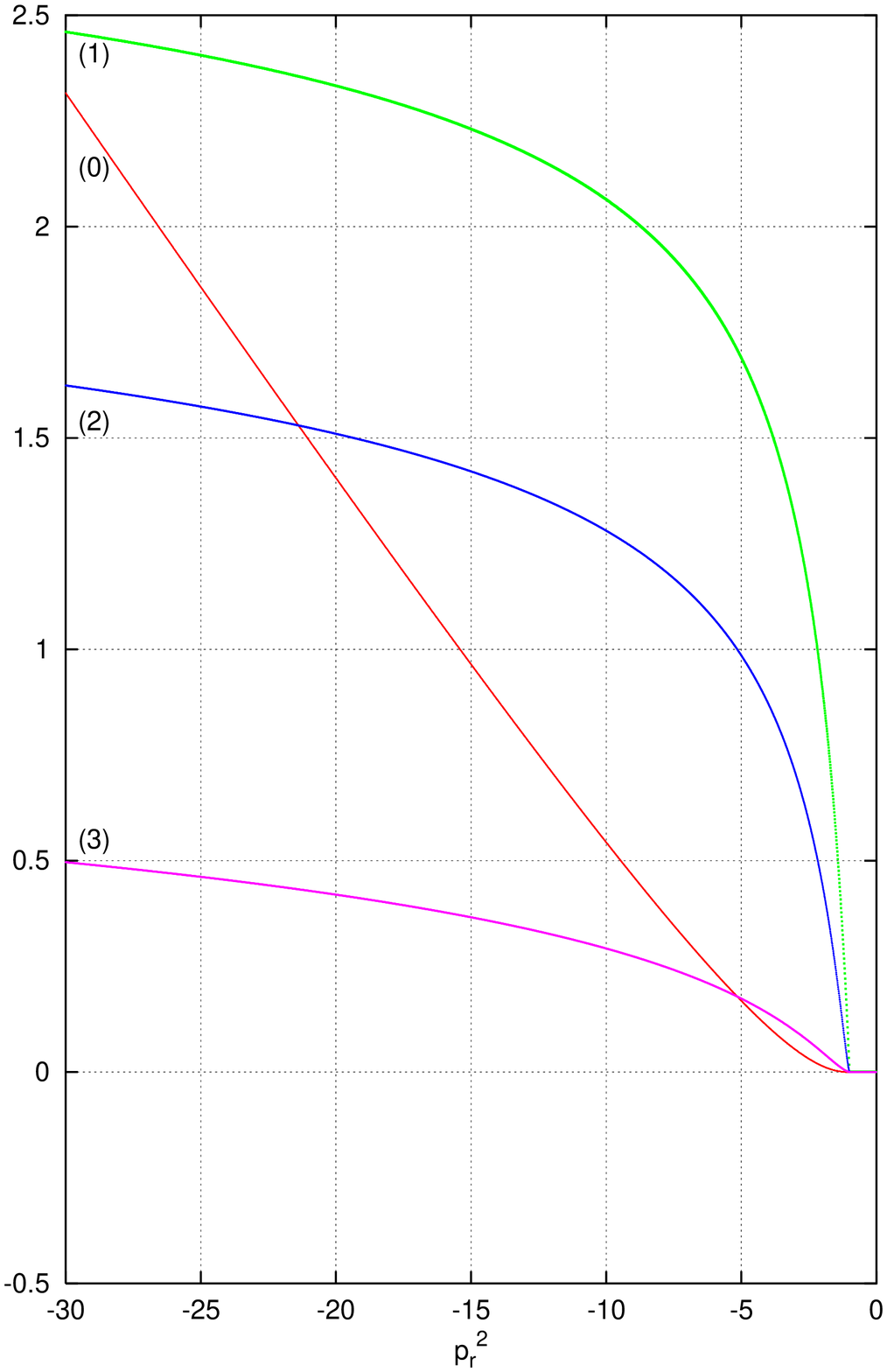,width=\textwidth,height=10cm}
\end{center}
\caption{Plots of 
\(\Im F^{(0)}_{0,r}\) (labeled as (0)) and \(\Im F^{(0)}_{i}\) 
(labeled as ($i$)) as a function of \(p_r^2\) for \(m_1 \ = \ 1, \ 
m_2\ = \ 9, \ m_3\ =\ 200 \) and  \(\mu = m_1+m_2+m_3 \). }
\label{fig:f4}
\end{figure}

The imaginary parts of all the functions plotted in figures
{\ref{fig:f3}} and {\ref{fig:f4}} exhibit no complicated structure
and their asymptotic behaviors can be simply deduced from \Eq{an1}
and \Eq{an2}. 
Observe that, due to our definition of $p^2_r$, 
the proper analytic continuation for time-like $p$, 
hence negative $ p^2_r = - |p^2_r| $, is obtained by giving a positive 
infinitesimal imaginary part to $ - p^2_r $, so that 
$\log(p_r^2) \to \log(p_r^2-i \epsilon) = \log|p_r^2| -i \pi$ 
(at variance with the default option of FORTRAN compilers).  
 
\section{The numerical method.}

For the numerical solution of the system of differential equations 
we use fourth-order Runge-Kutta method \cita{PTVF}. 
The method starts from the known values of the solutions in an 
initial point, 
then it calculates the values of the solutions in a nearby point at 
distance $\Delta$ with an expansion in $\Delta$ based on the differential 
equations, omitting terms of order $\Delta^5$. 
Repeating the procedure along a path of length $L$ in $N$ steps, so that
the step is of length $\Delta = L/N$, a relative error of approximately 
$N \Delta^5 = L^5/N^4$ is accumulated, and the requested accuracy 
is obtained by a suitable choice of $L$ and $N$. 
This method is known for its robustness, and indeed it works 
quite well in our case allowing us to obtain a relative accuracy of 
\(10^{-10}-10^{-12}\) (the FORTRAN program is written in double precision)
within reasonable CPU time (see discussion at end of this section).
More sophisticated methods exist and could be implemented, but the 
simplicity of the used one has the advantage of a better control on the 
accuracy.

To obtain the four MI related to the general massive 2-loop sunrise 
self-mass diagram we use the system of four linear differential equations 
given in \Eq{a6}. 
For the necessary initial conditions we use the values of the MI in the 
special points, where the differential equations simplify, allowing the 
analytic calculation of the MI; but as also the coefficients of 
the derivatives vanish there, to start the numerical evaluation from 
that points the values of the first derivatives must be provided as well. 
We use for that purpose the analytic values presented 
in \cita{CCLR1,CCR1,CCR2}. 
Starting from \(p_r^2 = 0\), numerical instabilities arise 
when approaching any of the pseudo-thresholds; therefore to obtain 
the value of the MI in the proximity of the threshold 
\(p_{th,r}^2 =-1\) or of the pseudo-thresholds 
$p_{ps1,r}^2, p_{ps2,r}^2, p_{ps3,r}^2$, we take 
the threshold or pseudo-thresholds themselves as the starting points 
of the numerical evaluation. 
Further, for very large values of \(p_r^2\) it is more convenient to start 
from the value \(x=\frac{1}{p_r^2}=0\), 
so that the asymptotic expansion at large \(p_r^2\) is also needed.  

We use the Runge-Kutta method in the complex plain of \(p_r^2\); 
the initial condition guarantees as in the real case 
the uniqueness of the solution, provided that we do not cross 
the cut \cite{I}, which extends in the present case from the threshold 
($p_{th,r}^2= -1$) to \(-\infty \) along the real axis.
As already remarked, due to use of Euclidean variables in \Eq{1}, 
the proper sign of the imaginary part of the solutions is obtained with 
a path laying in the lower half complex plane of \(p_r^2\). 
Using \(p_r^2\) as a complex variable has the advantage that the
initial conditions at \(p_r^2=0\) can be used to obtain the  
solutions everywhere, 
with a path which does not approach too much pseudo-thresholds and 
threshold. 
This feature is relevant, as the method reaches much faster the required 
accuracy, when starting from \(p_r^2=0\), rather than from anyone of the 
other special points  -- even if the result is of course independent 
of the chosen path.

The program is organized as an independent subroutine, whose 
arguments are the input values of the masses \(m_i\), of \(p^2\) 
(which is real) and for the required accuracies, 
and the output values of the MI with their errors.
Actually the input accuracies refer 
separately to the real part, the imaginary part and the absolute value 
of the functions. However the accuracy of the imaginary part is controlled 
by the program only when not vanishing, i.e. for \(p^2_r < -1\), but 
its value is in any case an indication of the precision of the result.
The program starts usually from the initial conditions at \(p^2_r =0\). 
However, if \(|p^2_r| >900\), it starts from initial conditions at 
\(x\equiv 1/p^2_r = 0\), while if \(|p^2_r-p^2_s| < 0.001\), where 
\(p^2_s\) is the threshold or any of the pseudo-thresholds, it 
starts from the initial conditions at \(p^2_s\).  

When starting from \( {p^2_r} = 0\), the fourth-order Runge-Kutta 
method is applied to the system of equations \Eq{a6} and 
the required final value of $ p^2_r $ is reached following  
the {\it rectangular} path in the complex plane: 
(0,0),(0,-0.1),(\( {p^2_r}\),-0.1),(\( {p^2_r}\),0).
When pointing to time-like values \( {p^2_r} < 0\), this path 
has the merit of avoiding the numerically troublesome points of the 
threshold and pseudo-thresholds. The same path is also used 
for reaching space-like values \( {p^2_r} > 0\), 
although no special points occur along on the real axis, 
as it turns out that the requested precision is usually reached much 
faster along a complex path than along the real one. 

The fourth-order Runge-Kutta for the system of equations \Eq{a6} 
is also used when the starting point of \( {p^2_r} \) is one of the 
pseudo-thresholds $p_{ps1,r}^2, p_{ps2,r}^2, p_{ps3,r}^2$, 
along a similar {\it rectangular} path: 
(\(p^2_{ps,r},0),(p^2_{ps,r},-0.1),({p^2_r},-0.1),( {p^2_r},0)\).

When starting from \( {p^2_r}=-1\) (the threshold) the same system of 
equations cannot be used, as the first derivatives of the  
master integrals \(F_i, \ i=1,2,3\) are infinite at that point.   
Instead, we introduce new functions $ F^{th}_{0,r},$, 
$ F^{th}_{i} $ through the definitions  
\begin{eqnarray}
 F^{(0)}_{0,r} &=& -\frac{\pi}{32} \sqrt{m_{1,r} m_{2,r} m_{3,r}}
 \  x^2_{th} \log(x_{th}) + F^{th}_{0,r} \nonumber \\
 F^{(0)}_{i} &=& 
 \frac{\pi}{16} \frac{\sqrt{m_{1,r} m_{2,r} m_{3,r}}} {m_{i,r}} 
 \  x_{th} \log(x_{th}) + F^{th}_{i} \ , \ i=1,2,3 \ , 
\labbel{alg1}
\end{eqnarray} 
\noindent
where \(x_{th}= p^2_r+(m_{1,r}+m_{2,r}+m_{3,r})^2 = p^2_r+1 \), 
we generate algebraically the system of differential equations which 
they satisfy, and then we solve numerically that new system within 
the program.  
We do not report here the new system of equations as one can easily obtain it 
from the \Eq{a6} using \Eq{alg1}. 
The function \(F^{(0)}_{0,r}\) 
does not have the same problems of the \(F^{(0)}_i,\ i=1,2,3\), but the 
subtraction is performed anyway to simplify the equations.
The required value of $p^2_{r}$ is then reached along the 
{\it triangular} path in the complex plane  
(-1,0), ( $(p^2_{r}-1)/{2}\;$,-0.01),( ${p^2_r}\;$,0).

In the asymptotic region \(|p^2_r| >900\) we perform the change of variables
\( p^2_r \rightarrow x\equiv 1/p^2_r \), then we subtract from the MI the 
terms not vanishing at \(x=0\) (the original MI are indeed divergent at 
\(x \to 0 \) ) and finally we write a system of equations for the subtracted 
MI $F_{0,r}^{as}$ and $F^{as}_{i}, \ i=1,2,3$ defined as  
\begin{eqnarray}
 F^{(0)}_{0,r} &=& \frac{p^2_r}{32} \left(\log(p^2_r)-\frac{13}{4}\right)
 +\frac{1}{32} \log^2(p^2_r)\sum_{i=1}^3 m_{i,r}^2
 -\frac{1}{16} \log(p^2_r)\sum_{i=1}^3 m_{i,r}^2\log(m_{i,r}^2)\nonumber \\
  &-&\frac{1}{32}\sum_{i=1}^3 m_{i,r}^2
  \left(5-6\log(m_{i,r}^2)+\log^2(m_{i,r}^2)\right) + F_{0,r}^{as}\nonumber \\
 F^{(0)}_{i} &=& -\frac{1}{32} \log^2(p^2_r)
  +\frac{1}{16} \log(p^2_r)\left(\log(m_{i,r}^2)+1\right)\nonumber \\
  &+&\frac{1}{32}\left(-1-4\log(m_{i,r}^2)+\log^2(m_{i,r}^2)\right)
  + F_{i}^{as} \ .
\labbel{alg2}
\end{eqnarray} 

Again the new system of equations for $F_{0,r}^{as}$, $F_{i}^{as}$ 
can be obtained in a simple way substituting \Eq{alg2} into \Eq{a6}. 
The numerical solution is then obtained in the variable $x$ 
along the complex {\it triangular} path 
(0,0),( ${x}/{2}$ ,-0.01),(\( x\),0).  

The errors assigned to the final results of the Runge-Kutta method 
are estimated by comparing them to the results obtained with a number of 
steps 10 times smaller then for the final results. The difference of 
the two results is taken as the estimate of the absolute error.
To account for the cumulated rounding error, 
we estimate the relative error in a $N$ step calculation as 
$\sqrt{N} \times 10^{-15}$, 
(as the program works in double precision), 
and then take the cumulated rounding 
error as the relative error times the value of the result. 
We finally take the sum of the absolute error and the cumulated rounding 
error as an indication of the error in the result.  

The initial number of steps $ N_i $ for \(|p^2_r|<1\) is taken to be 
$ N_i = 2/{min(accuracies)} $ (where $min(accuracies)$ is the smallest 
of the accuracies required in calling the routine) 
and $ N_i = 2|p^2_r|/{min(accuracies)} $ for \(|p^2_r|\ge 1\),
but is set to $ N_i=20$ if the number comes out smaller then 20.
If the required precision is not reached the number of steps
is increased by a factor 4, the system is solved once more
and the procedure for estimating the error is repeated.
For high required accuracy it might happen that the estimated
error grows when the number of steps is increased (because of an 
accumulation of the rounding errors, etc.). In that case the program
gives out the best result (i.e. the one with the smallest error). 
It may also happen, in the case the accuracy obtained in a given step 
is almost equal to the required one, that in the next step the accuracy 
obtained is much higher then the required one.  

\begin{table}
\begin{center}
\begin{tabular}{cllll}
 $p^2_r$ & $\Re F^{(0)}_{0,r}$ &      $\Im F^{(0)}_{0,r}$ 
& \kern-3pt$\Re F^{(0)}_1$ & \kern-7pt$\Im F^{(0)}_1$     \\ \hline  
\kern-11pt{\small -1000.} &\kern-9pt{\small -113.349786296(3)}
                          &\kern-5pt{\small   96.922241476(2)}
                          &\kern-7pt{\small   -0.4811355157(1)}
                          &\kern-7pt{\small    3.255217395(1) }\\
{\small -30.} &\kern+2pt{\small -0.230629539580(2)}
              &{\small  2.31604333072(1)}
              &\kern-3pt{\small 2.86686026427(1)}
              &\kern-7pt{\small 2.46075767262(1) }\\
{\small -15.} &\kern+6pt{\small 0.044413679180(8)}
              &{\small 0.964565210432(6)}
              &\kern-3pt{\small 3.47247970401(2)}
              &\kern-7pt{\small 2.23096593908(2)}\\
{\small -1.5} &\kern+2pt{\small -0.2536003902785(5)}
              &{\small  0.0032295545873(1)}
              &\kern-3pt{\small 5.03180993338(1)} 
              &\kern-7pt{\small 0.564979990606(8)}\\
\hline
\end{tabular}
\end{center}
\caption{ The benchmark values of \(F^{(0)}_{0,r}\) and \(F^{(0)}_1\)
 for masses \(m_1 = 1 \), \(m_2 = 9 \), \(m_3 = 200 \) and 
 \(\mu = m_1+m_2+m_3 \). }
\label{tab:tab1}
\end{table}

Typical running times on PC with Intel Pentium III (1GHz) CPU 
are the following:
for required accuracy of \(10^{-7}\) a fraction of a second for
\(|p^2_r|\simeq 0.2\), 2 seconds for \(|p^2_r|\simeq 2\) and 8 seconds
for \(|p^2_r|\simeq 30\); for required accuracy of \(10^{-11}\) 
20 seconds for \(|p^2_r|\simeq 0.2\), 3.5 minutes for \(|p^2_r|\simeq 2\) 
and 59 minutes for \(|p^2_r|\simeq 30\).  

\begin{table}
\begin{center}
\begin{tabular}{cllll}
 $p^2_r$ & $\Re F^{(0)}_2$ &            $\Im F^{(0)}_2$ 
& \kern-4pt$\Re F^{(0)}_3$ & \kern-8.1pt$\Im F^{(0)}_3$ \\ \hline  
\kern-11pt{\small -1000.} &\kern-3pt{\small -1.4689531571(1)}
                                   &{\small  2.393163661(1)}
                          &\kern-8pt{\small -0.81238274708(3)}
                          &\kern-8pt{\small  1.17931522036(6) }\\
{\small-30.} &{\small 0.880149099634(4)}
             &{\small 1.624323639603(8)} 
             &\kern-4pt{\small   0.1097269441157(5)}
             &\kern-8pt{\small 0.496397637240(4)} \\
{\small-15.}&{\small 1.26943734837(1)}
            &{\small 1.42094925579(1)}
            &\kern-4pt{\small   0.195266696001(4)}
            &\kern-8pt{\small 0.366097572324(2)}\\
{\small-1.5}&{\small 2.066516140875(4)}
            &{\small 0.239828506599(7)}
            &\kern-4pt{\small   0.2185459121886(5)}
            &\kern-8pt{\small 0.0195952353591(3)} \\
\hline
\end{tabular}
\end{center}
\caption{ The benchmark values of \(F^{(0)}_2\) and \(F^{(0)}_3\)
 for masses \(m_1 = 1 \), \(m_2 = 9 \), \(m_3 = 200 \) and 
 \(\mu = m_1+m_2+m_3 \). }
\label{tab:tab2}
\end{table}

The program is available from authors upon request and 
we report in Tables {\ref{tab:tab1}},{\ref{tab:tab2}} and {\ref{tab:tab3}}
a few results, which can serve as a benchmark. 
The reported results were all obtained asking the accuracies to be 
\(10^{-11}\).
In Table {\ref{tab:tab3}}
the values at \(p_r^2=0\), at threshold (\(p_r^2=-1\)) and at three 
pseudo-thresholds (\(p_r^2 = p_{ps1,r}^2, p_{ps2,r}^2, p_{ps3,r}^2 \)) 
are calculated from the known analytical results \cite{CCLR1,BDU,CCR1,CCR2} 
incorporated into the program, 
therefore no error is indicated (the imaginary parts vanish for 
those values of \(p_r^2\)). 

\begin{table}
\begin{center}
\begin{tabular}{cllll}
  \kern-5pt$p^2_r$   & \kern-7pt$F^{(0)}_{0,r}$ &\kern-7pt$F^{(0)}_1$ 
& $F^{(0)}_2$ & \kern-5pt $F^{(0)}_3$ \\ \hline  
\kern-5pt{\small-1} &\kern-7pt{\small -0.279454902855371}
                    &\kern-7pt{\small  4.83093725524177}
                    &\kern-5pt{\small  1.89253423642110}
                    &\kern-5pt{\small  0.185424043556224}\\
\kern-5pt{\small-0.99} &\kern-7pt{\small -0.2798928396415(5)}
                       &\kern-7pt{\small  4.80520291678(1)}
                       &\kern-5pt{\small  1.884686031133(5)}
                       &\kern-5pt{\small  0.1846128537367(4)}\\
\kern-5pt{\small\(p_{ps3,r}^2\)} &\kern-7pt{\small -0.280281633048667}
                                 &\kern-7pt{\small  4.78848044412341}
                                 &\kern-5pt{\small  1.87869903011051}
                                 &\kern-5pt{\small  0.183940826371101}\\
\kern-5pt{\small-0.9} &\kern-7pt{\small -0.2836780811878(5)}
                      &\kern-7pt{\small  4.68669935479(1)} 
                      &\kern-5pt{\small  1.836380342548(8)}
                      &\kern-5pt{\small  0.1786103121374(3)}\\ 
\kern-5pt{\small\(p_{ps2,r}^2\)} &\kern-7pt{\small -0.286233415451605}
                                 &\kern-7pt{\small  4.62888926299134}
                                 &\kern-5pt{\small  1.80974413459626}
                                 &\kern-5pt{\small  0.174898795219002}\\
\kern-5pt{\small-0.825} &\kern-7pt{\small -0.2866587055221(5)}
                        &\kern-7pt{\small  4.620058473554(8)} 
                        &\kern-5pt{\small  1.805561498336(6)} 
                        &\kern-5pt{\small  0.1742958493924(4)} \\
\kern-5pt{\small\(p_{ps1,r}^2\)} &\kern-7pt{\small -0.286906928933491}
                                 &\kern-7pt{\small  4.61498769104262}
                                 &\kern-5pt{\small  1.80314761916276}
                                 &\kern-5pt{\small  0.173945546345814}\\
\kern-5pt{\small-0.8} &\kern-7pt{\small -0.2876221116285(5)}
                      &\kern-7pt{\small  4.60069487253(1)} 
                      &\kern-5pt{\small  1.796297920597(6)}
                      &\kern-5pt{\small  0.1729424293568(4)}\\
\kern-5pt{\small-0.1} &\kern-7pt{\small -0.3101507246241(3)}
                      &\kern-7pt{\small  4.261426874519(4)} 
                      &\kern-5pt{\small  1.619837556072(1)}
                      &\kern-5pt{\small  0.1432933391084(2)} \\
\kern-5pt{\small 0} &\kern-7pt{\small -0.312816604092084}
                    &\kern-7pt{\small  4.22788075922252} 
                    &\kern-5pt{\small  1.60134292154365}
                    &\kern-5pt{\small  0.139821925866842} \\
\kern-5pt{\small 1.0} &\kern-7pt{\small -0.3340476235037(6)}
                      &\kern-7pt{\small  3.970691522343(6)} 
                      &\kern-5pt{\small  1.455289136414(2)}
                      &\kern-5pt{\small  0.1103513887593(2)} \\
\kern-5pt{\small 30.0}&\kern-7pt{\small \kern+4pt 0.323213333716(3)}
                      &\kern-7pt{\small 2.33632892937(2)} 
                      &\kern-5pt{\small 0.425519391740(4)}
                      &\kern-5pt{\small \kern-4pt -0.178170159306(2)}\\
\kern-9pt{\small 1000.}&\kern-7pt{\small \kern-7pt 115.539777092(3)}
                       &\kern-7pt{\small \kern-4pt  -0.8050335305(2)} 
                       &\kern-5pt{\small \kern-4pt  -1.7888035846(2)}
                       &\kern-5pt{\small \kern-4pt  -1.11980894482(3)}\\
\hline
\end{tabular}
\end{center}
\caption{ The benchmark values of \(F^{(0)}_{0,r}\), \(F^{(0)}_1\),
 \(F^{(0)}_2\) and \(F^{(0)}_3\)
 for masses \(m_1 = 1 \), \(m_2 = 9 \), \(m_3 = 200 \) and
 \(\mu = m_1+m_2+m_3 \). }
\label{tab:tab3}
\end{table}

\section{Tests and comparisons.}

Several checks were done in the past \cite{BDU,CCR1,CCR2} to verify 
that the analytical 
expansions of the master integrals in the special points, used 
within the numerical program,  
satisfy the differential equations and agree with the 
results existing in the literature. 

A remarkable feature of the extension of the RK-method 
to the complex plane is that it provides some natural self-consistency
checks of the algorithm implementation.
Starting from a special point and moving to a chosen value of $p^2$
with different paths in the $p^2$-complex-plane, the values obtained 
for the master integrals should agree inside the errors of the 
method discussed previously. One has however to remember that paths 
chosen in the upper and lower half-plane, respect to the real axis, 
give opposite sign to the imaginary part of the master integrals 
for time-like values of the external invariant above threshold, 
$ p^2_r < -1 $.  
An even more complete test is to reach the same value of $p^2$ 
starting from different special points, hence following 
different paths, and compare the values of the master integrals at $p^2$ 
obtained along the various paths.  
If the values coincide inside the assigned errors the consistency 
between the differential equations, the expansions in the special points 
used as initial values, the implementation of the RK-method and the 
algorithm for estimating the errors are cross-tested in a rather 
effective way.  

We have performed several of the mentioned checks in the different 
regions of $p^2$ obtaining the requested agreement.

The only published precise numerical results for the general massive 
case (all different non-zero mass values) are presented in 
\cite{BBBS,Passarino}, in the form of a combination of the general 
massive case with massless cases, to cancel the pole singularities in $(n-4)$.
In our notation that combination is
\begin{eqnarray} 
T_{123N}(p^2,m_1^2,m_2^2,m_3^2) = -16 \bigl[ 
&+& F_0^{(0)}(m_1^2,m_2^2,m_3^2,p^2) -F_0^{(0)}(m_1^2,0,m_3^2,p^2) \nonumber \\
&-&F_0^{(0)}(0,m_2^2,m_3^2,p^2)     +F_0^{(0)}(0,0,m_3^2,p^2) \bigr] \ ,
\labbel{T123N} \end{eqnarray} 
which has also the property of being independent of $\mu$; 
the overall factor $(-16)$ accounts for the different definition 
of the master integral in \Eq{1} and our $p^2$ corresponds to $(-p^2)$
in \cite{BBBS} and to $s$ in \cite{Passarino}. 

To obtain the values for $F_0^{(0)}(0,0,m_3^2,p^2)$ we use the 
analytic formula presented in \cite{CCLR1}, while for the values of
$F_0^{(0)}(m_1^2,m_2^2,m_3^2,p^2)$,
$F_0^{(0)}(m_1^2,0,m_3^2,p^2)$ and $F_0^{(0)}(0,m_2^2,m_3^2,p^2)$ 
we use the present program. Although the value zero for the masses is 
not allowed, we have checked that the limit can be in practice reached
numerically.
Comparing the results obtained for the mass values from $10^{-6}$ to 
$10^{-9}$, we can estimate the error coming from having a mass not 
exactly zero, by taking the difference 
between the results obtained with the two smallest values used for
the mass to be set to zero. 
As the error due to zero mass limit is sometimes comparable with 
the error due to the RK-method, we sum the two errors 
for each of the considered functions.  
The final error of $T_{123N}$ is the sum of the errors assigned 
by the algorithm to each of the contributing functions in \Eq{T123N}. 
The larger errors (or less efficiency in calculations) come from the 
zero mass contributions, for which an approach based entirely on 
analytical expressions is in preparation \cite{CGZ}.  
Furthermore the choice of equal values for two or even all the three masses, 
reduces the number of the independent equations in the system of 
differential equations, generating potential numerical problems, 
although less serious than those for the zero mass.

In \cite{BBBS} the values for $T_{123N}(p^2,m_1^2,m_2^2,m_3^2)$ are 
presented, for different sets of the masses $m_1, m_2, m_3$, and for 
the two regions of small $|p^2|<(m_1+m_2+m_3)^2$ and large 
$|p^2|>(m_1+m_2+m_3)^2$ in their Table 1 and 2 respectively.

We repeat in Table \ref{tab:tab4} for the same values of the masses and $p^2$ 
the results of Table 1 of \cite{BBBS} for the multiple series (first 
entry), pushed to a large number of terms in some cases, and our results 
(second entry). The results are in excellent agreement.
\renewcommand{\baselinestretch}{0.9}
\begin{table}
\begin{center}
\begin{tabular}{rrr|rl|rl}
  $m_1$ & $m_2$ & $m_3$ & $p^2$ & $T_{123N}$  & $p^2$ & $T_{123N}$
  \\ \hline  
 {\small  3} & {\small  3} & {\small  10} & 
 {\small -9} & {\small -7.3129877443} & 
 {\small  9} & {\small -6.93244055931}\\
 {\small   } & {\small } & {\small } & 
 {\small   } & {\small -7.3129877442(26)} & 
 {\small   } & {\small -6.93244055924(50)}\\
\hline
 {\small   2} & {\small  3} & {\small  10} & 
 {\small -20} & {\small -4.1493850173 } & 
 {\small  20} & {\small -3.63591843327 }\\
 {\small    } & {\small } & {\small } & 
 {\small    } & {\small -4.1493850171(18) } & 
 {\small    } & {\small -3.63591843320(95) }\\
\hline
 {\small  2} & {\small  2} & {\small  10} & 
 {\small -25} & {\small -2.3353847298  } & 
 {\small  25} & {\small -1.9428452190 }\\
 {\small    } & {\small } & {\small } & 
 {\small    } & {\small -2.3353847298(14) } & 
 {\small    } & {\small -1.9428452191(10) }\\
\hline
 {\small  1} & {\small  2} & {\small  10} & 
 {\small -30} & {\small -0.8117674738 } & 
 {\small  30} & {\small -0.6306847352 }\\
 {\small    } & {\small } & {\small } & 
 {\small    } & {\small -0.8117674738(15) } & 
 {\small    } & {\small -0.6306847353(12) }\\
\hline
 {\small  1} & {\small  1} & {\small  10} & 
 {\small -49} & {\small -0.3167501084 } & 
 {\small  49} & {\small -0.1950338472  }\\
 {\small    } & {\small } & {\small } & 
 {\small    } & {\small -0.3167501085(24) } & 
 {\small    } & {\small -0.1950338472(21) }\\
\hline
 {\small  3} & {\small  4} & {\small  15} & 
 {\small -50} & {\small -7.9471022759  } & 
 {\small  50} & {\small -6.8270303849 }\\
 {\small    } & {\small } & {\small } & 
 {\small    } & {\small -7.9471022760(33) } & 
 {\small    } & {\small -6.8270303852(19) }\\
\hline
 {\small  3} & {\small  4} & {\small  20} & 
 {\small -100} & {\small -6.0171476156 } & 
 {\small  100} & {\small -4.9485063889 }\\
 {\small     } & {\small } & {\small } & 
 {\small     } & {\small -6.0171476159(59)  } & 
 {\small     } & {\small -4.9485063897(42)  }\\
\hline
 {\small  3} & {\small  4} & {\small  20} & 
 {\small -150} & {\small -6.3903568683  } & 
 {\small  150} & {\small -4.7506023184 }\\
 {\small     } & {\small } & {\small } & 
 {\small     } & {\small -6.3903568686(73) } & 
 {\small     } & {\small -4.7506023184(66)  }\\
\hline
 {\small  5} & {\small  5} & {\small  25} & 
 {\small -150} & {\small -14.5339444977  } & 
 {\small  150} & {\small -1.21816923644 }\\
 {\small     } & {\small } & {\small } & 
 {\small     } & {\small -14.5339444982(87)  } & 
 {\small     } & {\small -1.21816923648(68)  }\\
\hline
 {\small  5} & {\small  5} & {\small  25} & 
 {\small -200} & {\small -15.0523063012  } & 
 {\small  200} & {\small -11.8790613597  }\\
 {\small     } & {\small } & {\small } & 
 {\small     } & {\small -15.0523063010(95) } & 
 {\small     } & {\small -11.8790613597(83)  }\\
\hline
\end{tabular}
\end{center}
\caption{ Comparison for small \( |p^2| < (m_1+m_2+m_3)^2\). 
In each box the first entry is the value of the multiple series 
of Table 1 of \cite{BBBS}, the second entry is our result 
(the error in the last digits is enclosed in parenthesis). 
Our value of \(p^2\) corresponds  to \(-p^2\) in \cite{BBBS}. 
  }
\label{tab:tab4}
\end{table}
\renewcommand{\baselinestretch}{1.1}

Also in Table 7 of \cite{Passarino} the values for the same combination 
$T_{123N}$ (there called $S_c$) are presented for small $s$, equal to our 
\(p^2\), $|p^2| < (m_1+m_2+m_3)^2$ and for $m_1=10, m_2=20, m_3=100$. 
They are repeated here in Table \ref{tab:tab7}, where in each box the 
first entry comes from the multiple series of \cite{BBBS} with a large 
number of terms, the second entry is the present result, the  
third entry is from the numerical integration of \cite{Passarino}.
Again we have excellent agreement with the multiple series of 
\cite{BBBS},
while the accuracy of the numerical integration of \cite{Passarino} 
is within a few $ppm$, inside the relative $10^{-5}$ precision declared 
there.  

\renewcommand{\baselinestretch}{0.9}
\begin{table}
\begin{center}
\begin{tabular}{rl|rl|rl}
  $p^2$ & $T_{123N}$  & $p^2$ & $T_{123N}$& $p^2$ & $T_{123N}$
  \\ \hline  
   {\small  -1}  & {\small -70.6856984 } 
 & {\small  -25} & {\small -70.75620346 } 
 & {\small  -81} & {\small -70.92141286 }\\
   {\small     } & {\small -70.6856977(39) } 
 & {\small     } & {\small -70.75620352(11) } 
 & {\small     } & {\small -70.92141291(70)  } \\
   {\small     } & {\small -70.686011 } 
 & {\small     } & {\small -70.756299 } 
 & {\small     } & {\small -70.921481 } \\
\hline
   {\small  1}  & {\small -70.6798310 } 
 & {\small  25} & {\small -70.609519049 } 
 & {\small  81} & {\small -70.446146678 }\\
   {\small    } & {\small -70.6798305(21) } 
 & {\small    } & {\small -70.609519051(97) } 
 & {\small    } & {\small -70.446146655(35) } \\
  {\small     } & {\small -70.680106 } 
 & {\small    } & {\small -70.609231} 
 & {\small    } & {\small -70.446044 } \\
\hline
\end{tabular}
\end{center}
\caption{ Comparison for small \( |p^2| < (m_1+m_2+m_3)^2\) and for 
$m_1=10, m_2=20, m_3=100$.
In each box the first entry is the multiple series 
value of \cite{BBBS}, the 
second entry is our result (the error on the last digits 
is in parenthesis) the third entry is from the numerical integration 
in Table 7 of \cite{Passarino}.
Our value of \(p^2\) corresponds  to \(-p^2\) in \cite{BBBS} 
and to $s$ in \cite{Passarino}. 
  }
\label{tab:tab7}
\end{table}
\renewcommand{\baselinestretch}{1.1}

In Table \ref{tab:tab5} we report the results of Table 2 of \cite{BBBS}
for the combination $T_{123N}$,  for large and negative \(p^2\) 
(i.e. $|p^2| > (m_1+m_2+m_3)^2$, the value of \(p^2\) here corresponds to 
\(-p^2\) in \cite{BBBS}), so that this time there is also an imaginary part. 
In each box the first entry comes from the multiple series of \cite{BBBS}, 
with a large number of terms, the second entry is the present result.
Again we have excellent agreement with the multiple series 
of \cite{BBBS} 
in most of the cases, in few cases there is a deviation of two times the 
assigned error, that we attribute to our procedure of approaching the 
zero mass. 
In the seventh box we assign an error also to the multiple series, because, 
although each sum is taken up to 70 terms, the results are not yet stable. 
The assigned error is the difference with the sums taken up to 60 terms. 
We attribute the difficulty to the chosen value of $p^2=-150$, which is 
too near to the threshold value $-(3+4+5)^2=-144$.

\renewcommand{\baselinestretch}{0.9}
\begin{table}
\begin{center}
\begin{tabular}{rrrr|ll}
  $m_1$ & $m_2$ & $m_3$ & $p^2$ & $\Re T_{123N}$   & $\Im T_{123N}$
  \\ \hline  
 {\small  2} & {\small  3} & {\small  2} &  {\small -80} & 
 {\small 0.587432001  }     & {\small -11.262835755 }\\
 {\small   } & {\small   } & {\small   } & {\small     } & 
 {\small 0.587431990(41)  } & {\small -11.262835744(24)  }\\
\hline
 {\small  3} & {\small  3} & {\small  3} & 
 {\small -100} & {\small -1.28284949 }     & {\small -20.899600723 }\\
 {\small   } & {\small } & {\small } & 
 {\small     } & {\small -1.28284943(16) } & {\small -20.899600741(10) }\\
\hline
 {\small  2} & {\small  3} & {\small  4} & {\small -100} & 
 {\small -0.3286481685  }    & {\small -11.84587606309 }\\
 {\small  } & {\small } & {\small } &  {\small     } & 
 {\small -0.3286481687(16) } & {\small -11.84587606302(60) }\\
\hline
 {\small  3} & {\small  4} & {\small  4} &  {\small -150} & 
 {\small -1.26795173  }    & {\small -26.491194705 }\\
 {\small  } & {\small } & {\small } &  {\small     } & 
 {\small -1.26795173(21) } & {\small -26.491194694(41) }\\
\hline
 {\small  2} & {\small  4} & {\small  3} & {\small -150} & 
 {\small 1.5662482672 }     & {\small -11.9689438774 }\\
 {\small  } & {\small } & {\small } & {\small     } & 
 {\small 1.5662482670(14) } & {\small -11.9689438778(12)  }\\
\hline
 {\small  3} & {\small  3} & {\small  4} & {\small -150} & 
 {\small 0.9865824304 }     & {\small -16.10213970663 }\\
 {\small  } & {\small } & {\small } & {\small     } & 
 {\small 0.9865824312(19) } & {\small -16.10213970687(77) }\\
\hline
 {\small  3} & {\small  4} & {\small  5} & {\small -150} & 
 {\small -4.7638745(12) }    & {\small -29.601246304(26) }\\
 {\small  } & {\small } & {\small } & {\small } & 
 {\small -4.7638748416(32) } & {\small -29.60124631204(68)  }\\
\hline
 {\small  2} & {\small  3} & {\small  4} & {\small -200} & 
 {\small 1.6960823345 }     & {\small -6.02417248918  }\\
 {\small  } & {\small } & {\small } & {\small } & 
 {\small 1.6960823350(17) } & {\small -6.02417248906(81)  }\\
\hline
 {\small  3} & {\small  4} & {\small  4} & {\small -200} & 
 {\small 1.86355967 }     & {\small -20.39852237 }\\
 {\small  } & {\small } & {\small } & {\small } & 
 {\small 1.86355979(34) } & {\small -20.39852240(12)  }\\
\hline
 {\small  4} & {\small  4} & {\small  4} & {\small -250} & 
 {\small 2.64395201 }     & {\small -27.60904430 }\\
 {\small  } & {\small } & {\small } & {\small } & 
 {\small 2.64395222(31) } & {\small -27.60904439(18)  }\\
\hline
\end{tabular}
\end{center}
\caption{ Comparison for \(p^2\) large $(|p^2| > (m_1+m_2+m_3)^2)$ 
and negative. 
In each box first entry is the multiple series value of Table 2 
of \cite{BBBS}, the 
second entry is our result (the error on the last digits is 
in parenthesis). 
Our \(p^2\) corresponds  to \(-p^2\) in \cite{BBBS}. 
  }
\label{tab:tab5}
\end{table}
\renewcommand{\baselinestretch}{1.1}

In Table \ref{tab:tab6} we report the results of Table 2 of \cite{BBBS}
for the combination $T_{123N}$, for large and positive \(p^2\) 
(i.e. $|p^2| > (m_1+m_2+m_3)^2$; our \(p^2\) corresponds to 
\(-p^2\) in \cite{BBBS}).
In each box the first entry comes from the multiple series of \cite{BBBS}, 
with a large number of terms, the second entry is the present result.
Also here we have excellent agreement with the multiple series 
of \cite{BBBS} 
in most of the cases, in few cases there is a deviation about two times the 
assigned error, that we attribute to our procedure of approaching the 
zero mass. 

\renewcommand{\baselinestretch}{0.9}
\begin{table}
\begin{center}
\begin{tabular}{rrr|rl|rrr|rl}
  $m_1$ & $m_2$ & $m_3$ & $p^2$ & $T_{123N}$  & $m_1$ & $m_2$ & $m_3$ 
 &$p^2$ & $T_{123N}$
  \\ \hline  
 {\small  2} & {\small  2} & {\small  2} & {\small 50} &
 {\small -3.728125558 } &
 {\small  2} & {\small  4} & {\small  3} & {\small 100} &
 {\small -7.1836810855 }\\
 {\small  } & {\small } & {\small } & {\small } &
 {\small -3.728125610(46) } & 
 {\small  } & {\small } & {\small } & {\small  } &
 {\small -7.1836810854(54) }\\
\hline
 {\small  3} & {\small  3} & {\small  3} & {\small 100} & 
 {\small -8.79126989 } & 
 {\small  3} & {\small  4} & {\small  3} & {\small 120} & 
 {\small -12.430997190}\\
 {\small  } & {\small } & {\small } & {\small } &
 {\small -8.79127000(11) } & 
 {\small  } & {\small } & {\small } & {\small  } &
 {\small -12.430997250(30) }\\
\hline
 {\small  3} & {\small  3} & {\small  4} & {\small 150} & 
 {\small -7.0830520665} & 
 {\small  3} & {\small  4} & {\small  4} & {\small 150} & 
 {\small -10.85647158 }\\
 {\small  } & {\small } & {\small } & {\small } &
 {\small -7.0830520661(28) } & 
 {\small  } & {\small } & {\small } & {\small  } &
 {\small -10.85647158(26) }\\
\hline
 {\small  3} & {\small  4} & {\small  3} & {\small 150} & 
 {\small -11.361931056 } & 
 {\small  3} & {\small  4} & {\small  4} & {\small  200} & 
 {\small -9.64611359 }\\
 {\small  } & {\small } & {\small } & {\small } &
 {\small -11.361931121(31) } & 
 {\small  } & {\small } & {\small } & {\small  } &
 {\small -9.64611360(29) }\\
\hline
 {\small  2} & {\small  3} & {\small  4} & {\small 200} & 
 {\small -3.3018636831 } & 
 {\small  4} & {\small  4} & {\small  4} & {\small  250} & 
 {\small -13.57188440 }\\
 {\small  } & {\small } & {\small } & {\small } &
 {\small -3.3018636830(21) } & 
 {\small  } & {\small } & {\small } & {\small  } &
 {\small -13.57188463(21) }\\
\hline
\end{tabular}
\end{center}
\caption{ Comparison for \(p^2\) large $(|p^2| > (m_1+m_2+m_3)^2)$ 
and positive. 
In each box first entry is the multiple series value of 
Table 2 of \cite{BBBS}, the 
second entry is our result (the error on the last digits is 
in parenthesis). Our \(p^2\) corresponds  to \(-p^2\) in \cite{BBBS}. 
  }
\label{tab:tab6}
\end{table}
\renewcommand{\baselinestretch}{1.1}

\section{Conclusions.}

We propose to solve numerically, by means of the Runge-Kutta method 
extended to the complex plane, the system of the differential equations 
satisfied by the MI related of the diagrams, which due to the large number 
of occurring parameters cannot be calculated analytically. 

We apply the method to the study of the simplest non trivial diagram, 
the general massive 2-loop sunrise self-mass, which is 
already exhibiting a number of intriguing analytic properties.  
We obtain, for all the allowed values of the parameters and of the 
external invariant $p^2$, very accurate values of 
the MI within reasonable CPU time, in good agreement with the 
results already present in the literature.  
 
The method can be naturally extended to the other self-mass diagrams 
of the same order, like the 2-loop 4-propagator self-mass diagram 
for which the differential equation is already  
known \cite{CCLR2}.

The extension to higher order self-mass diagrams will only increase 
linearly the number of the MI and of the differential equations in 
the system, while the growing of the number of parameters is not a 
problem at all.
Also the extension to diagrams with three or more external legs, 
which means multi variable cases, can be easily envisaged.

The true difficulty of the method is the need of initial conditions 
for starting the numerical solution of the differential equations; 
clearly the initial conditions have to be provided 
by an independent method. Of special values are, in this respect, 
the special points (such as $0, \infty,$ thresholds and pseudothresholds) 
where an analytic calculation is easier and sometimes possible, as in the 
case discussed in this paper. When the special points are used, also the 
first derivative of the MI have to be provided as an independent input to 
the Runge-Kutta approach, but that is analytically a relatively simpler task, 
amounting to an iteration of the expansions provided by 
the differential equations.  

\vskip 0.4 cm

{\bf Acknowledgments.}
We thank Sandro Rambaldi for his invaluable advise on the use of
Runge-Kutta method to solve differential equations.

One of us (HC) is grateful to the Bologna Section of INFN and to the Department
of Physics of the Bologna University for support and kind hospitality.

\vfill \eject 
\appendix

\section{Corrections to some analytic formulae of \cite{CCR1,CCR2}}
\label{app:bugs}

 We report here for completeness the correct form of the formulae,
 which are wrongly reported in our previous publications \cite{CCR1,CCR2} 
 and are used here to obtain the numerical values of the MI at 
 pseudo-thresholds and threshold.
 
 In section 5 of \cite{CCR1} there are three misprints. In Eq.(41)
 the factor \(\frac{1}{16}\) in front of the integral should be missing.
 In the first line of Eq.(44) \(\log(y)\) should read \(\log(y_S)\)
 and in Eq.(47) there is a missing factor 4 in front of \({\cal I}_2\).

 In \cite{CCR2} there is one misprint in Eq.(45): the second line from
 the end should be of the opposite sign (\('-' \to '+'\)).
 In Eq.(31) of \cite{CCR2} the expression for ${\cal I}_3(m_1,m_2,m_3)$ 
 should be symmetric in all the masses, so it becomes 

 \begin{eqnarray}
 {\cal I}_3(m_1,m_2,m_3)=\tilde{\cal I}_3(m_1,m_2,m_3)
 +\tilde{\cal I}_3(m_1,m_1,m_2)-\tilde{\cal I}_3(m_2,m_1,m_1) \ ,
  \labbel{app1}
 \end{eqnarray}
with
 \begin{eqnarray}
&&{\tilde{\cal I}_3(m_1,m_2,m_3)}=
 \sqrt{m_1 m_2} \int dm_3 \frac{1}{\sqrt{m_3(m_1+m_2+m_3)}}
  \left[\frac{\log\left(\frac{m_3}{m_1}\right)}{m_3+m_1}
     +\frac{\log\left(\frac{m_3}{m_2}\right)}{m_3+m_2}\right]= \nonumber \\
 && \kern-12pt i\Biggl\{\log\left(\frac{m_1+m_2}{4m_1}\right)
   \left[  \log(t-t_1)-\log(t-t_2)\right]
      +\log\left(\frac{m_1+m_2}{4m_2}\right)
     \left[\log(t+t_2)-\log(t+t_1)\right]  \nonumber \\
  &&+\log(t-t_1)\left[2\log(1-t_1)-\log(t_1)\right]
  -\log(t-t_2)\left[2\log(1-t_2)-\log(t_2)\right]\nonumber \\
  &&-\log(t+t_1)\left[2\log(1+t_1)-\log(-t_1)\right]
  +\log(t+t_2)\left[2\log(1+t_2)-\log(-t_2)\right]\nonumber \\
  &&-2\ \hbox{Li}_2\left(\frac{t-t_1}{1-t_1}\right)
    +2\ \hbox{Li}_2\left(\frac{t-t_2}{1-t_2}\right)
    +\hbox{Li}_2\left(-\frac{t-t_1}{t_1}\right)
    -\hbox{Li}_2\left(-\frac{t-t_2}{t_2}\right)\nonumber \\
  &&+2\ \hbox{Li}_2\left(\frac{t+t_1}{1+t_1}\right)
    -2\ \hbox{Li}_2\left(\frac{t+t_2}{1+t_2}\right)
    -\hbox{Li}_2\left(\frac{t+t_1}{t_1}\right)
    +\hbox{Li}_2\left(\frac{t+t_2}{t_2}\right) \Biggr\}
    \ , \labbel{EX12} \end{eqnarray} 

where the expressions of the logarithms account now properly for their 
imaginary part. 
Consequently the Eq.(42) of \cite{CCR2} should be replaced by

 \begin{eqnarray}
 b-\frac{K}{32} = \pi\left(-\frac{1}{32}
    +\frac{5}{32}\log(2)\right)
   +\frac{1}{8}\hbox{Cl}_2\left(\frac{\pi}{2}\right)
\ . \labbel{EX22} \end{eqnarray} 

\vfill \eject 
\def\NP{{\sl Nucl. Phys.}} 
\def\PL{{\sl Phys. Lett.}} 
\def\PR{{\sl Phys. Rev.}} 
\def\PRL{{\sl Phys. Rev. Lett.}} 
\def\NC{{\sl Nuovo Cim.}}
\def\APP{{\sl Acta Phys. Pol.}}
\def\ZP{{\sl Z. Phys.}}
\def\MPL{{\sl Mod. Phys. Lett.}} 
\def\EPJ{{\sl Eur. Phys. J.}} 
\def\IJMP{{\sl Int. J. Mod. Phys.}}

\end{document}